\documentstyle[12pt,epsf]{article}
\makeatletter
\def\@cite#1#2{$^{\hbox{\scriptsize{#1\if@tempswa , #2\fi})}}$}
\makeatother

\begin{document}
\title{Self-Annealing Dynamics in a Multistable System}
\author{ Tsuyoshi Hondou \\ 
 Graduate School of Information Sciences\\
 Tohoku University,
 Sendai 980-77\\
}
\date{}
\maketitle
\begin{abstract}
 A new type of dynamical behavior of a multistable system is reported.
 We found that a simple non-equilibrium system can reduce
its effective temperature  
 autonomously  
at a global minimum if the residual frustration at a 
global minimum is small enough,
 which  
highlights an unexpected feature of non-equilibrium multistable
 systems.
\end{abstract}


How can one escape from a labyrinth?
The question is important not only for a game player but also
for a researcher for multistable systems. 
Evolution of complex systems sometimes obey an optimization process
of a kind of "energy" function. 
Since the landscape of the energy is in general multistable, 
the harm of trap at local minima prevents the system from reaching a 
global minimum.
A conventional way to escape from local minima is to make thermal fluctuation
in such
systems, and the relaxation process under thermal fluctuation has been 
extensively
studied.

However, there is a dilemma between a barrier crossing probability
in a multistable potential\cite{RMP,Hanggi} and a stationary 
probability distribution of
the system:
High temperature fluctuation makes a barrier crossing between basins
of multistable potential easy:
 But, unfortunately, 
high temperature fluctuation also makes a mean energy over a probability
distribution increase.
Simulated annealing method\cite{Kirk}, which is a strategy to lead a system to 
a global minimum by gradually decreasing the temperature of the thermal
fluctuation, was 
introduced to avoid the dilemma 
and was successfully applied for several fields such as 
image 
restoration\cite{GG}, 
protein folding\cite{PF2}, neural 
networks\cite{NN} and so on. 
 However, the method seems somewhat 
unnatural and inconvenient for physical processes
because of the two reasons: 1) one must "control" 
the 
temperature of a system gradually, because the convergence to a 
global minimum is guaranteed when one spends infinite time to decrease the
temperature; 2) the system never stops even if the system 
passes the global minimum state as long as the temperature is finite.
This implies that the dilemma cannot be solved essentially
even by the "simulated annealing" method.

 Although the simulated annealing method is a kind of relaxation
process, one should remember the fact that the method is 
based on the {\em equilibrium}
statistical mechanics because it uses Boltzmann distribution,
which is realized for the system with detailed balance, 
in its process,
 Therefore, we conjectured
that the dilemma might be solved in a non-equilibrium condition.
The conjecture is partially motivated by the Levinthal's paradox
that a protein is folded from its initial structure
much quicker than the exhaustive sampling\cite{Levinthal}.
 We report in this paper a preliminary example of a solution of the dilemma 
by numerical simulation of a simple model.


 One finds that the "energy" of a multistable system is 
 generally
 composed of plural competitive constraints or interactions
 each of which is relatively simple, and the competition makes frustration
in a system.
 The global minimum is, thereby, the state where such constraints are 
mostly
satisfied: the frustration is expected to be the minimum there. 
 An interesting example of this competitive dynamics can be found in  
on-line learning process of neural networks.
 Radons {\em et al.} reported \cite{Radons}
 that an effective temperature
of parameter fluctuations caused by a probabilistic successive
pattern input in a learning process by backpropagation can decrease 
autonomously at a
global minimum under condition that all  
constraints (patterns)
can be perfectly satisfied at a global minimum\cite{Heskes}.
The condition is called as "perfectly trainable", which is approximately
satisfied for the network with a sufficiently large number of 
neurons\cite{Funahashi}.
 The autonomous decrease of fluctuation was also found in the learning
process of chaotic time series by a conventional feedforward neural 
network\cite{Prog}
and in the learning by the neural network with coupled 
oscillators\cite{Inoue2}.
Recent study of on-line learning by neural networks indicates that 
 non-thermal fluctuation due to the successive change of 
the local potential (error)
 is important not only for escape from local minima but also for an 
acceleration of a relaxation process even {\em without} local minima\cite{PRE}, 
which needs a clear understanding of the effect of non-thermal
fluctuations due to fluctuating potentials.

 The dynamics which autonomously changes the effective temperature is quite interesting from the view point of
relaxation process in non-equilibrium condition.
 However, the discussions above were limited for 
learning by neural networks, and dynamical analysis of a "perfectly
trainable" system 
 has not been carried out due to their high-dimensionality and their
complex structures.
 To investigate the autonomous, annealing-like dynamics clearly, 
we find a minimal model which satisfies the following conditions:
1) Each partial potential (constraint) term, $V^{i}(x)$, is expressed in a positive definite,
differentiable 
function
so that a zero value state corresponds to a frustration-free state; 
2) Global potential, $V(x)$, is a linear sum of the constraints;
3) The dynamics obeys a plain relaxation (gradient descent) process of a 
 dissipative particle under time-dependent potentials;
4) A system has small residual frustration in a global minimum state;
5) A functional form of the partial potential is selected to make the 
residual frustration of a global minimum small in order to demonstrate a 
typical self-annealing dynamics;
6) The amplitudes of the partial potentials fluctuate in
order to make non-equilibrium fluctuation. 

 A dynamics which satisfies these conditions is realized:
\begin{equation}
  x_{n+1} = x_{n} -\Delta t \frac{ \partial V^{p_{n}}(x)}{\partial x}|_{x=x_{n}}
,
\label{dynamics}
\end{equation}
 where a global potential is $V(x) =\frac{1}{N}\sum_
{i=1}^{N} V^{i}(x)$; a partial potential with index $i$ is 
 $V^{i}(x) = (1-
\cos(a_{i} x + \delta_{i}))/2$ $[$for $ -10 \le x \le 10$, 
otherwise $V^{i}(x)
 = \infty]$;  the coefficients $a_{i}$ and $\delta_{i}$ 
are arbitrary constants;
 $p_{n}$ is an index of the partial 
potential at discrete time, $n$, which is chosen randomly from 
the partial potential indices,
\{1,2,3, $\cdots$, $N$\}; and $\Delta t$ is a constant.
 It should be noticed that local minima of the global potential
are not attributed to one of the local minima of the partial potential,
but are attributed to the result of 
interference of all partial potentials, which is 
consistent with general multistable systems. The multistability of the
partial potential is not unnatural at least for the learning 
by neural networks,
because the number of available solutions is generally plural in the systems.
 The "energy" landscape discussed here is multistable and
has a global minimum near $x=0$ when the phase shift coefficient,
 $|\delta_{i}|$,
 is small enough.
The pre-factor $1/N$ in the potential is only for a normalization
purpose.

 It should be noticed that the dynamics of eq.\ref{dynamics} 
corresponds to an extreme example of the self-annealing dynamics that 
the amplitudes of 
partial potentials fluctuate independently: 
only one potential is alive each time step in this extreme case. 
However, the result is not essentially altered 
when plural partial potentials are alive, whereas the
non-equilibrium effect is weakened in this case. 
It should also be noticed that the central result is the same  for the 
continuous time dynamics with slowly varying amplitudes of
 partial potentials with time scale of
$\Delta t$. 
The dynamics coincides with a plain relaxation process of a "quenched" 
global potential in  
the limit $\Delta t \rightarrow 0$, because the system "feels" an average 
potential. Therefore, the finiteness of the parameter,
$\Delta t$, is essential for the "non-equilibrium" dynamics.

 We find that the present system has a "self-annealing"\cite{Inoue} ability
 that fluctuation in phase space decreases at a global minimum {\em
autonomously}
 if the 
amplitudes of constraints (partial potentials) fluctuate 
as in eq.(1).
 First we show the most ideal case that the system is in a frustration-free
state
at a global minimum, which is realized by setting all the coefficients, 
$\delta_{i}$, 
to be zero. 
Fig.1 shows that the motion of the system started in a basin of a local
minimum stops suddenly when the system reaches a 
global minimum state, which is a typical example of the
 "self-annealing dynamics".
 It should be emphasized that the
frustration is not reduced much at a local minimum, 
in contrast with the global minimum
state. 
 This dynamics is qualitatively different from a 
conventional dynamics (Langevin 
equation) subject to  
thermal noise (Fig.2):
\begin{equation}
x_{n+1}= x_{n} -\Delta t \frac{ \partial V(x)}{\partial x}|_{x=x_{n}} 
+ \xi_{n},
\end{equation}
where $\xi$ is a probabilistic random noise with zero mean.
 This observation of the self-annealing dynamics is understandable 
because the perfectly
 global minimum condition ($V=0$) implies
that each of partial potentials is simultaneously
 minimized at the global minimum ($V^{i}=0$),  
 while the local minimum condition 
($\frac{\partial V}{\partial x} =0 $) 
does not imply the stability at a local minimum for all partial potentials 
( $\exists i: 
\frac{\partial 
V^{i}}{\partial x} \neq 0$; or 
$\frac{\partial^{2} V^{i}} {\partial x^{2}} < 0$). 
 
 The discussion above may seem too ideal, since the perfect minimum 
condition, 
$V(x)=0$, must be assumed at a global minimum state. Next, 
we discuss 
the case that the perfect minimum condition is not
satisfied even in a global 
minimum state (Fig.3a). The situation is realized when some of the 
partial potentials, $V^{i}$,
are not minimized at the global minimum, due to the finite values of
$\delta_{i}$. As shown in Fig.3b,
numerical simulation shows that the system never 
stops at a global minimum, however,
the system can reduce its 
activity (fluctuation of the system) when the system reaches a global minimum
and can be stabilized enough to stay in the global minimum.
The stability of a system at the global minimum state 
in the present potential 
is guaranteed if the 
following inequality is satisfied: 
$ \Delta t \cdot \max_{i} |a_{i}|
 < \min_{i,j} |(\pi - \delta_{i})/ 
a_{i} -  \delta_{j} / a_{j} | $, which is derived from the worst condition that
a particle cannot be transferred to a next basin of any partial potential by
an overshooting of the largest potential gradient of partial potentials. 
 The inequality roughly explains the stability at a global minimum:
 increase of $\Delta t$, $|a_{i}|$ and $|\delta_{i}|$ breaks the stability at a 
global minimum. Because an increase of $|\delta_{i}|$ from zero increases the 
residual frustration at a global minimum,
the "self-annealing" dynamics 
works less effectively if residual frustration at a global minimum is
not small. 
 
We showed in this paper that an autonomous, annealing-like dynamics
("self-annealing dynamics")
is possible in a non-equilibrium multistable system where
the residual frustration at a global minimum is small enough:
The self-annealing dynamics has a unique
property as to the convergence toward a global minimum, in contrast with
simulated
annealing: 
the
system under self-annealing dynamics decreases its effective
temperature {\em by itself}
suddenly when the system reaches a
global minimum.	
Non-equilibrium multistable systems can be
qualitatively different from equilibrium multistable systems
of which fluctuation is characterized by thermal temperature. 
In general multistable systems, the global minimum is the state, by definition, where 
each interaction is mostly satisfied in average; which implies that the
frustration of the system
 is expected to be mostly solved in a global minimum state\cite{JDB}.
Therefore, the present simulation suggests that 
other non-equilibrium system with different forms of interactions (constraints)
may have a possibility 
to show the self-annealing dynamics.

In many body systems
 with exchange interactions, the global 
potential is conventionally composed of pairs of
elements with coupling coefficients, where a spin glass system is a prototype.
Conventional solid state material is usually assumed that the exchange
interaction
is fixed as known in "quenched"  systems.
In such quenched systems, the self-annealing dynamics does not 
appear. 
However, we
conjecture that some many body systems such as biological systems
 which are far from equilibrium 
condition might show 
the self-annealing dynamics, since the exchange interaction may 
fluctuate in such systems.
 We also expect that 
the self-annealing dynamics
might be useful for 
  optimization problems where 
the energy (or error) function of the problem is made
by the composition of plural constraints: successive evolution of the system
under the presentation of a partial constraint can 
 produce the self-annealing dynamics. 
 The studies of such realizations
are under way. 

 The author would like to thank M.Sano, Y.Hayakawa, K.Tokita,
 M.Nakao, M.Yamamoto
 and T.Chawanya
for stimulating discussions and Y.Sawada for fruitful discussions and
helpful comments. This work is supported in part by the Japanese Grant-in-Aid
for Science Research Fund from the Ministry of Education, 
Science and Culture (No.
07740315).

\clearpage

\begin{figure}[h]
\caption{
Temporal evolution of the self-annealing dynamics in the most ideal case.
$a$) Landscape of a total potential energy, $V(x)$, where
 the number, $N$, of partial potentials
is 10. The coefficients, $a_{i}$, are chosen as: \{  1, 1.38, 2.75, 1.96, 1.27, 1.64, 1.42, 2.29, 2.47, 2.72\} for demonstration. 
The coefficients, $\delta_{i}$, are chosen as zero in order to stabilize 
the global minimum state perfectly. 
$b$) Typical trajectories of a phase space and a total energy.
}
\end{figure}

\begin{figure}[h]
\caption{
Temporal evolution of the conventional Langevin equation subject to 
random noise (eq.(2)), where $\xi_{n}$ is a uniform random number, $r_{n}$,
 $|r_{n}|< 
0.5$. Typical trajectories of a phase space and a total energy are 
shown. The potentials used are the same as in Fig.1. 
}
\end{figure}

\begin{figure}[h]
\caption{
Temporal evolution of the self-annealing dynamics in the less ideal case.
$a$) Landscape of a total potential energy.
 The number, $N$, of partial potentials and the coefficients, $a_{i}$,
 are chosen as in Fig.1.
 The coefficients, $\delta_{i}$, are chosen as: \{-0.7, 0.2, -0.2, 0.1, -0.6,
0.3, -0.5, 0.1, -0.3, 0.2\} respectively corresponding to $\{a_{i}\}$ in
Fig.1. 
$b$) Typical trajectories of a phase space and a total energy.
}
\end{figure}


\clearpage

\begin{thebibliography}{99}
\bibitem{RMP}
P.H\"anggi, P.Talker and M.Borkovec, Rev.Mod.Phys. {\bf 62} (1990), 251
and the references therein.
\bibitem{Hanggi}
G.F.Fleming and  P.H\"anggi, (eds.), {\em Activated Barrier Crossing} 
(World Scientific, Singapore, 1993). 
\bibitem{Kirk}
S.Kirkpatrick, C.D.Gelatt and M.P.Vecchi, 
{\em Science} {\bf 220} (1983), 671.
\bibitem{GG}
S.Geman, and D.Geman, IEEE Trans. PAMI {\bf 6} (1984), 721.
\bibitem{PF2}
V.I.Abkevich, A.M.Gutin and E.Shakhnovich, 
J.Chem.Phys. {\bf 101} (1994), 6052.
\bibitem{NN}
 E.Aarts and J.Korst, {\em Simulated annealing and Boltzmann machines} (Wiley, Chichester, 1989).
\bibitem{Levinthal}
 C.Levinthal,  J.Chim.Phys. {\bf 65} (1968), 44.
\bibitem{Radons}
 G.Radons, H.G.Schuster and D.Werner, in {\em Parallel Processing in Neural
 Systems and Computers} (ed. R.Eckmiller)  p.261
 (Elsevier Science Pub., Amsterdam, 1990).
\bibitem{Heskes}
 A more general analysis of the dynamics in a feedforward neural network 
is found in, T.M.Heskes and B.Kappen, Phys.Rev {\bf A 44} (1991), 2718.
\bibitem{Funahashi}
In feedforward neural networks with no less than three layers,
 any continuous function can be approximated with any precision.
A proof appears in,  K.Funahashi, Neural Networks {\bf 2} (1989), 183.
\bibitem{Prog}
See Fig.4 of the paper:
T.Hondou and Y.Sawada,  Prog.Theor.Phys. {\bf 91}  (1994), 397.
\bibitem{Inoue2}
M.Inoue, A.Tanaka and K.Nakamoto, Prog.Theor.Phys. {\bf 93} (1995), 845.
\bibitem{PRE}
T.Hondou, M.Yamamoto, Y.Sawada and Y.Hayakawa, Phys.Rev.{\bf E 53} (1996), 
4217.
\bibitem{Inoue}
 The same word, self-annealing, was used to describe the effect of
the fluctuation
caused by a chaotic sequence on 
an optimization problem by a chaos neural network:
M.Inoue and A.Nagayoshi, Prog.Theor.Phys. {\bf 88} (1992), 769. 

\bibitem{JDB}
 A related concept as the present study called the "principle of minimal 
frustration" can also be found in a protein folding problem;
see J.D.Bryngelson and P.G.Wolynes, Proc.Natl.Acad.Sci. (U.S.A.) {\bf 84}
 (1987), 7524.
\end{thebibliography}
\end{document}